\documentstyle[prl,aps]{revtex}
\begin{document}
\draft
\twocolumn[\hsize\textwidth\columnwidth\hsize\csname@twocolumnfalse\endcsname
\preprint{}
\title{Superconductivity in Ferromagnetic 
RuSr$_2$GdCu$_2$O$_8$ }
\author{Jian-Xin Zhu,$^{(1)}$ C. S. Ting,$^{(1)}$ and 
C. W. Chu$^{(1,2)}$
}
\address{ 
$^{(1)}$Texas Center for Superconductivity and Department of Physics, 
University of Houston, Houston, Texas 77204\\
$^{(2)}$Lawrence Berkeley National Laboratory, 1 Cyclotron Road, 
Berkeley, CA 94720}
\maketitle
\begin{abstract}
The phase diagram of temperature versus exchange field is obtained within 
a BCS model for $d$-wave  
superconductivity in CuO$_2$ layers which is coupled to 
ferromagnetic RuO$_2$ layers
in RuSr$_2$GdCu$_2$O$_8$.  
It is found that the Fulde-Ferrell-Larkin-Ovchinnikov 
state is very sensitive to the band filling factor. For strong exchange 
field, we point out that superconductivity could only 
exist in the interfaces between ferromagnetic domains. 
The magnetization curve is calculated and its comparison with 
experiment is discussed.
We also propose the measuring of tunneling conductance near a single 
unitary impurity to detect 
the strength of the exchange interaction. 
\end{abstract} 
]

\narrowtext
The problem of coexistence of superconductivity (SC) and 
ferromagnetism (FM) 
has attracted keen interest since the original works of 
Ginzburg~\cite{Ginzburg57} and Matthias {\em et al.}~\cite{MSC58}. 
It was shown that singlet SC and FM are mutually 
exclusive and the SC can also be 
strongly suppressed by magnetic impurities. 
The competition between SC and FM were observed in the ternary 
compounds, HoMo$_6$S$_8$, HoMo$_6$Se$_8$, and ErRh$_4$B$_4$~\cite{MF82}. 
But true microscopic coexistence was found 
only over a narrow temperature region when FM sets in and modifies 
itself to a spiral or domain-like structure.
The recent discovery of SC ($T_c=16$-$47$ K) in the ferromagnetic 
($T_{M}=132$ K) ruthenate-cuprate layered compound RuSr$_2$GdCu$_2$O$_8$ 
(Ru-1212)~\cite{Tallon99,Bernhard99,Pringle99,Hadjiev99,Chu99,Bernhard00,Lynn00} 
renewed the interest in the issue of how 
SC and FM negotiate to coexist. Recent band structure calculation 
performed by Pickett {\em et al.}~\cite{PWS99} showed that the 
exchange splitting in the CuO$_2$ layer is small 
($\Delta_{\mbox{\small exc}}   
=25\;\mbox{meV}$) compared to $\sim 1\;\mbox{eV}$ in 
the RuO$_2$ 
layer but is larger enough that the superconducting state in the CuO$_2$ 
layer may be of the 
Fulde-Ferrell-Larkin-Ovchinnikov (FFLO) type~\cite{FF64,YS98} or finite 
momentum pairing state.  
Whether there actually exists such a new pairing state even in early found 
magnetic superconductors remains to be controversial~\cite{NLG81}. 
Indeed, until now no superconductor has been 
discovered to be a finite-momentum pairing state. Also notice that, in 
earlier studies, the FFLO state was discussed by 
assuming a constant density of states (DOS), that is, the exchange 
field does not introduce structures in the electronic properties on the 
energy scales relevant to SC. However, this becomes not true 
when the DOS has singular structure.  Thus a further study 
on the existence of this state by including the energy dependence of DOS 
near the Fermi surface should be interesting. 
Moreover, it was supposed 
that the SC mainly occurs in the CuO$_2$ layers in 
Ru-1212, the identification of the pairing symmetry in this material is 
also of fundmental importance in view of the well-established $d$-wave 
pairing symmetry in high-$T_c$ cuprate superconductors, which have 
similar crystal structures. 

The purpose of this paper is three fold: 
(i) By assuming a $d$-wave pairing symmetry in a two-dimensional (2D) 
lattice model, we present a detailed study of the temperature-exchange 
field phase diagram to invetigate the sensitivity of the FFLO state 
by varying  the position of the Fermi energy within the tight-binding band;  
(ii) Arguing that the mixed state is intrinsic, we calculate the 
magnetization as a function of an applied magnetic field and compare it 
with experiment; (iii) We propose to measure the existence of the zero-energy 
peak (ZEP) and its splitting in the differential tunneling conductance 
near a single unitary impurity in the CuO$_2$ layer as a test of the 
$d$-wave pairing symmetry as well as the strength of the exchange field.    

Our model system is defined on a 2D lattice 
with pairing interaction taking place between two electrons on the 
nearest-neighbor sites, which in the mean field approximation leads to 
the Bogoliubov-de-Gennes equations
\begin{equation}
\sum_{j} \left( \begin{array}{cc} 
H_{ij,\sigma} & \Delta_{ij} \\
\Delta^{*}_{ij} & -H_{ij,\bar{\sigma}} 
\end{array} 
\right) 
\left( \begin{array}{c}
u_{j\sigma}^{n} \\
v_{j\bar{\sigma}}^{n}
\end{array} 
\right) 
=E_{n}\left( \begin{array}{c}
u_{i\sigma}^{n} \\
v_{i\bar{\sigma}}^{n} 
\end{array}
\right) \;,
\label{EQ:BdG}
\end{equation}
with the single particle Hamiltonian
$H_{ij,\sigma}=-t \delta_{i+\gamma,j} 
-\mu \delta_{ij} -\sigma h_{\mbox{\small exc}}\delta_{ 
ij}+U_{i}\delta_{ij}$, 
and the self-consistent condition
$\Delta_{ij}=\frac{V}{4} \sum_{n,\sigma} 
(u_{i\sigma}^{n}v_{j\bar{\sigma}}^{n*} 
+ u_{j\bar{\sigma}}^{n}v_{i\sigma}^{n*})\tanh (E_{n}/2k_{B}T) $.
Here $(u_{i\sigma},v_{i\bar{\sigma}})$ are the Bogoliubov 
quasiparticle amplitudes on the $i$-th site; $\gamma=\pm 
\hat{x},\pm\hat{y}$ represents the relative position of sites 
nearest-neighboring to the $i$-th site;  $t$ is the effective hopping 
integral between two nearest-neighbor sites within the CuO$_2$ plane; 
$\mu$ is the chemical potential; $h_{\mbox{\small 
exc}}=J(\langle S_{z}^{a}\rangle +\langle S_{z}^{b}\rangle)$ 
is the exchange field coming from the ordered spins in the two 
nearest-neighboring ferromagnetic RuO$_2$ layers; 
$U_{i}$ if any accounts for the scattering from the impurities; $V$ is 
the strength of the nearest-neighbor pairing interaction.   
Notice that the internal magnetic field on CuO$_2$ layers due to the 
magnetic moment on RuO$_2$ layers
is only several hundred gauss and the exchange interaction 
should be dominant in suppressing SC.   
We therefore defer the effect of the internal field on the orbital 
motion of paired electrons to the study of the magnetization. 
 
{\em Temperature--exchange-field phase diagram}.--- For a pure system in 
the absence of external magnetic field, the 
Bogoliubov quasiparticle amplitude can be generally written as 
$\left( \begin{array}{c}
u_{i\sigma}^{n}\\
v_{i\bar{\sigma}}^{n}
\end{array}\right)
=\frac{1}{\sqrt{N}} \left( \begin{array}{c} 
u_{{\bf k}+{\bf q},\sigma} e^{i({\bf k}+{\bf q}) \cdot {\bf R}_{i}} \\
v_{{\bf k}-{\bf q},\bar{\sigma}}  
e^{i({\bf k}-{\bf q}) \cdot {\bf R}_{i}}
\end{array} \right)$,
which yields the bond order parameter  
\begin{eqnarray}
\Delta_{ij}&=&\frac{V}{2N} e^{i{\bf q}\cdot ({\bf R}_{i}+{\bf R}_{j})} 
\sum_{\bf k} \cos[{\bf k}\cdot ({\bf R}_{j}-{\bf R}_{i})] 
\nonumber \\
&& \times [u_{{\bf 
k}+{\bf q},\uparrow} v_{{\bf k}-{\bf q},\downarrow}^{*}
+u_{{\bf 
k}+{\bf q},\downarrow} v_{{\bf k}-{\bf q},\uparrow}^{*}
] \tanh(E_{{\bf k},{\bf q}}/2T) \nonumber \\
&=&\Delta_{\delta}^{(0)}(i) e^{i{\bf q}\cdot ({\bf R}_{i}+{\bf R}_{j})} \;.
\label{EQ:OP1}
\end{eqnarray}
Here $N=N_x \times N_y$ is the number of two-dimensional lattice sites. 
Eq.~(\ref{EQ:OP1}) shows that the order parameter is not a constant in 
space manifesting the collective motion of paired electrons each with 
momentum ${\bf q}$. 
Using the definition $\Delta_{d}=(\Delta_{\hat{x}}^{(0)}
+\Delta_{-\hat{x}}^{(0)}-
\Delta_{\hat{y}}^{(0)}-\Delta_{\hat{y}}^{(0)})/4$, we find the  
equation determing the $d$-wave energy gap
\begin{equation}
1=\frac{V}{4N} \sum_{{\bf k},\sigma} \frac{(\cos k_x-\cos k_y)^{2}}
{E^{(0)}_{{\bf k},{\bf q}} } 
\tanh \frac{
E^{(1,\sigma)}_{{\bf k},{\bf q}} }{2T}
\;,
\label{EQ:OP}
\end{equation}
where 
$E^{(0)}_{{\bf k},{\bf q}}=\sqrt{
Z_{{\bf k},{\bf q},+}^{2}
+\vert \Delta_{\bf k}\vert^{2} } \;,
$
and 
$E^{(1,\sigma)}_{{\bf k},{\bf q}}=-\sigma h_{\mbox{\small exc}} 
+Z_{{\bf k},{\bf q},-}
+E^{(0)}_{{\bf k},{\bf q}}\;,
$
with $\xi_{\bf k}=-2t (\cos k_x +\cos k_y)-\mu$, 
$Z_{{\bf k},{\bf q},\pm}=(\xi_{{\bf k}+{\bf q}}\pm 
\xi_{{\bf k}-{\bf q}})/2$,
and $\Delta_{\bf 
k}=2\Delta_{d} (\cos k_x -\cos k_y)$.
Correspondingly, the free energy per lattice site is given by 
\begin{eqnarray}
{\cal F}&=&\frac{1}{2N}\sum_{{\bf k},\sigma} 
\{ (\xi_{{\bf k}+{\bf q}}-\sigma h_{\mbox{\small exc}}) \left( 
1+\frac{Z_{{\bf k},{\bf q},+}}{E_{{\bf k},{\bf q}}^{(0)}}\right) 
f_{{\bf k},{\bf q}}^{\sigma} 
\nonumber \\
&& + (\xi_{{\bf k}-{\bf q}}+\sigma h_{\mbox{\small exc}}) \left( 
1-\frac{Z_{{\bf k},{\bf q},+}}{E_{{\bf k},{\bf q}}^{(0)}}\right) 
[1-f_{{\bf k},{\bf q}}^{\sigma}] 
\nonumber \\
&&
+2T[(1-f_{{\bf k},{\bf q}}^{\sigma}) \ln (1-f_{{\bf k},{\bf 
q}}^{\sigma}) 
+ f_{{\bf k},{\bf q}}^{\sigma} \ln f_{{\bf k},{\bf q}}^{\sigma}]
\}
\nonumber \\
&& -\frac{4}{V}\vert \Delta_{d}\vert^{2} \;,
\label{EQ:FREE}
\end{eqnarray}
where $f_{{\bf k},{\bf q}}^{\sigma}
=f(E_{{\bf k},{\bf q}}^{1,\sigma})$ is the Fermi distribution function. 

To determine the phase boundary between the normal pairing (${\bf q}=0$)  
state and the normal state (spin polarized), 
one should compare the free energies of the superconducting 
state and normal state, using Eq.~(\ref{EQ:FREE}).  
In the presence of a fairly strong exchange interaction, the system might 
also go into the FFLO state in which all the 
Cooper pairs have a single non-vanishing (${\bf 
q}\neq 0$) center of mass momentum. 
This transition between the FFLO state and the normal state would 
be of the second order. To find the transition curve for the 
FFLO state and the normal state, we solve Eq.~(\ref{EQ:OP}) 
with $\Delta_d=0$ to find the maximum value of $h_{\mbox{\small exc}}$ by 
scanning through a whole set values of ${\bf q}$ at the same temperature $T$.
By repeating the same calculation at a different value of $T$, we then  
obtain the phase curve $h_{\mbox{\small exc}}$ as a function of $T$.   

To see the sensitivity of the FFLO state to the Fermi energy position in 
a 2D tight-binding band, we fix the pairing interaction as 
$V=2t$ and consider three typical values of the chemical 
potential, $\mu=-t$, $-0.5t$, and $-0.14t$, corresponding to the 
band-filling factor $\nu \approx 0.65$, 0.82, and 0.95 ($\nu=1$ is a 
half-filled band). For the above sets of parameters,  
the  maximum energy gap at zero temperature and $h_{\mbox{\small exc}}=0$ 
is found to be
$\Delta_{0}=4\Delta_d \approx 0.65t$,  $0.88t$, and 
$0.96t$, and  correspondingly, 
the transition temperature 
is $T_{c0}\approx 0.26t$, $0.37t$, and $0.40t$, respectively.
Figure~\ref{FIG:PHASE} plots the temperature--exchange-field phase 
diagram. Our calculation shows that the momentum 
${\bf q}$ corresponding to the maximum $h_{\mbox{\small exc}}$ is along the 
(10) and its equivalent direction in the whole 
temperature region because the energy gap reaches the maximum value along 
these directions so that the system can be more robust against the 
depairing effect from both the finite momentum and the 
exchange field.
 As shown in Fig.~\ref{FIG:PHASE}(a), when $\mu=-t$, 
the transition curve (solid line) is between the superconducting state (with 
${\bf q}=0$ and ${\bf q}\neq 0$) and the normal state, 
which shows that at low temperatures, when the exchange field is increased 
the system initially in the 
normal pairing state will enter the FFLO state $({\bf q}\neq 0)$ through 
the first-order transition  and then pass into the normal state by a 
second-order transition. Thus 
the FFLO state is a stable state at high exchange fields. The transition curve 
between the normal pairing state and the FFLO state is represented by dashed 
line. When the temperature is increased, two curves becomes closer and at 
$T=0.58T_{c0}$ (tri-critical point) they coincides with each other, 
where the transition begins to be of second order and the FFLO state 
merges naturally to the ${\bf q}=0$ normal pairing state.
This result is similar to that obtained within the continuum 
model using a constant DOS~\cite{FF64,YS98}. 
However, we find that the phase space for the existence of the FFLO state 
shrinks at $\mu=-0.5t$ as the band-filling factor shifts toward the half 
filling (Fig.~\ref{FIG:PHASE}(b)). In particular, near half filling when 
$\mu 
=-0.14t$, the FFLO state will be unstable and appear as a supercooling state.
Therefore, as shown in Fig.~\ref{FIG:PHASE}(c), only the 
transition between the normal pairing state and the normal state is 
physically acceptable. 
The novel feature comes from the 
influence of the exchange field on the DOS near the Fermi 
surface. For a 2D tight-binding band, the DOS at the energy 
$\mu=0$ has a singular 
point and it decays logarithmically as the energy goes 
away from zero. When  $\mu=-t$, which has 
been far away from the zero energy point, the DOS is flat  
for the energy near the Fermi surface. In this case, 
the splitting of the normal 
electron band by the exchange field has little effect on the change of 
the DOS and a constant DOS can be taken which 
corresponds to the approximation made in the continuum 
theory~\cite{FF64,YS98}. 
But as the chemical potential is close to the zero 
energy point, a little splitting of the energy band will cause a strong 
variation of the DOS near $\mu=0$, which makes the  
${\bf q}\neq 0$ pairing state unfavorable. 

For the Ru-1212, the band structure calculation estimated the Fermi 
wavevector as $\pi/a$. The filling factor should be near the 
half-filling ($\mu=0$). Then the existence of the FFLO state in the above 
system 
becomes unlikely according to Fig.~\ref{FIG:PHASE}(c). In the following 
discussion, we focus on 
the transition as between the normal pairing state and the normal state. 
Consider the case of $\nu=0.95$ (Fig.~\ref{FIG:PHASE}(c)). If the 
zero-field transition temperature is assumed to be $T_{c0}=90\;\mbox{K}$, 
$\Delta_{0}$ is about 18.7 meV. By fitting to the transition 
temperature $T_{s}\sim 36\;\mbox{K}$ as measured 
experimentally~\cite{Bernhard99}, the exchange field should be as small 
as 8.8 meV. Actually, $T_{c0}$ may be  as low as 60 K when Ru is 
replaced by other atoms (e.g., Cu)~\cite{Cao98}. Then the allowable 
$h_{\mbox{\small 
exc}}$ is only 5.1 meV to have $T_{s}\sim 36\;\mbox{K}$. 
Notice that the exchange field $h_{\mbox{\small exc}}$ from the band structure 
calculation is as large as 12.5 meV~\cite{PWS99} and the experimental 
filling is very close to $\nu=1$ (or $\mu=0$), we therefore conclude that, if 
$h_{\mbox{\small exc}}$ is indeed so 
large, SC can only exist in the CuO$_2$ layers between the 
ferromagnetic domains where the exchange interactions 
$h_{\mbox{\small 
exc}}=J(\langle S_{z}^{a}\rangle +\langle S_{z}^{b}\rangle)$
are small. 
It is important to emphasize that we have not ruled out the 
possibility of the FFLO state in the real systems because the estimated 
value of exchange interaction could in fact vary over a limited range 
depending on the method of obtaining it. In case that  the real exchange 
interaction is somewhat smaller than that estimated 
in Ref.~\cite{PWS99}, the bulk FFLO state would become a reality.

{\em Magnetization}.--- The 
bulk Meissner effect has not been observed in the superconducting state of 
Ru-1212~\cite{Chu99} until very recently~\cite{Bernhard00}.   
Furthermore, even the
absence of superconductivity has been reported in well characterized
Ru-1212 samples~\cite{Bauernfeind95}.
This discrepancy indicates  
the delicate balance between the superconducting and ferromagnetic 
interactions and the experimental result appears to depend critically on
the sample condition in a yet-to-be determined fashion. 
It is reported that the magnetic moment in each Ru atom is 1 Bohr 
magneton~\cite{Tallon99}. With structure parameter values, $a=3.8\;\AA$, 
and $c=11.4\;\AA$, the internal magnetic field 
$H_{int}=4\pi M_{sp}=1\mu_{B}/a^{2}c$ ($M_{sp}$ is the spontaneous 
magnetization) is estimated to be 
$707\;\mbox{G}$ in the CuO$_2$ layers which is larger than the 
first critical field $H_{c1}^{(0)}\sim 100\;\mbox{G}$ of a non-Ru layered 
cuprate superconductor with the comparable transition temperature,
i.e., $H_{c1}^{(0)}-4\pi M_{sp}<0$. The measured $H_{c1}$ is therefore zero 
and no bulk Meissner effect can be observed.   
Instead the superconductivity occurring in the CuO$_2$ 
layers has been driven into the mixed state.
The overall magnetization in the system consists of 
two parts, one from the spontaneous magnetization $M_{sp}$ of the 
ferromagnetic 
RuO$_2$ layers, the other $M_{ob}$ from the diamagnetic orbital 
contribution of 
the superconducting CuO$_2$ layers in the mixed state, i.e., 
$M=M_{sp}+M_{ob}$. 
When an external magnetic field $H_{ext}$ is applied, the effective magnetic 
field is $H=H_{ext}+H_{int}$. As an approximation, we work with the 
London equation for a square vortex lattice to find the magnetic induction 
$B\approx H-H_{c1} \ln 
(H_{c2}/B)/\ln \kappa$~\cite{FH69}, 
where $\kappa$ is the Ginzburg-Landau parameter 
and $H_{c2}=2\kappa^{2} H_{c1}$ is the upper critical field for the 
CuO$_2$ subsystem. In Fig.~\ref{FIG:MAG}, the total magnetization is 
plotted as a function of $H_{ext}$. As it is shown, the magnetization 
increases monotonically with the external magnetic field.
The observed monotonic behavior of $M$ by the 
experiment~\cite{Chu99} is shown in the inset of Fig.~\ref{FIG:MAG}.
Because our study is based upon the SC in a single ferromagnetic domain,  
the internal magnetic field must be greater than the intrinsic
first critical field $H_{c1}$ of the CuO$_2$ SC.
Nevertheless, when $H_{ext}=0$, our calculation gives an 
appreciable spontaneous magnetization which contradicts the experimentally 
measured zero magnetization~\cite{Chu99}.  This  
difference shows the existence of  
ferromagnetic domains in Ru-1212, where the average magnetization 
vanishes in the absence of an 
external magnetic field and the SC  
occurs in the interfacial regions between some of the ferromagnetic domains.
Note that the magnetization has also been recently studied~\cite{SF98} 
to analyze the superconducting properties of 
R$_{1.5}$Ce$_{0.5}$RuSr$_2$Cu$_2$O$_{10}$, 
where the Meissner effect was absent at temperature region 
$T_{d}<T<T_c$ ($T_c\sim 30\mbox{K}$ and $T_d\sim 20\mbox{K}$)
but present at $T<T_d$. To interpret this phenomenon, it was proposed 
that the ferromagnetism appears with a domain structure but the 
superconductivity is a bulk phase, which is different from our 
explanation for Ru-1212 systems, where neither Meissner effect (down to 
0.5 Gauss) nor detectable condensation energy was observed at temperature 
down to 2K~\cite{Chu99}.  Very recently, the detection of Meissner 
state at a field below 30 Oe was reported by Bernhard et 
al.~\cite{Bernhard00}. In the reported sample with a Meissner  
effect, the internal field was estimated to be only about 50-70 Oe (the   
lower critical field $H_{c1}$ of the nonmagnetic superconductor was estimated
to be of the order 80-120 Oe), in contrast to the previous reported 
$H_{int}$ about 200-700 Oe by the same group and others. 
We argue that the experimental result of Bernhard et al. might make sense 
if the superconductivity occurs in the inter-ferromagnetic domain region 
where $H_{int}$ is much reduced and the ratio $R$ between superconducting 
volume/sample volume is not too small.
This would not be inconsistent with the conclusion reached for 
Ref.~\cite{Chu99}, where
$R$ could be rather small so that the Meissner effect was not detected.
 
{\em Quasiparticle resonant state near a single unitary impurity in 
CuO$_2$ layers}.--- Since the ferromagnetic exchange field has pre-existed 
in the above system, it can affect the quasiparticle resonant states near 
a single impurity in the case of $d$-wave pairing symmetry. 
To address both issues, we solve 
Eq.~(\ref{EQ:BdG}) self-consistently using the exact diagonalization 
method~\cite{ZTH00} 
and calculate the local density of states (LDOS) 
$\rho_{i}=-\sum_{n,\sigma} [ \vert u_{i\sigma} \vert^{2} f^{\prime}(E-E_n)
+\vert v_{i\bar{\sigma}}\vert^{2} f^{\prime}(E+E_{n})]
$.
The calculation was made on $6\times 6$ supercells each with size 
$35a\times 35a$ by assuming a paramagnetic pairing state. 
The parameters are as follows: 
The single-site impurity strength $U_0=100t$, $\nu=0.95$, and $T=0.02t$.
From Eq.~(\ref{EQ:BdG}), we can see that 
the zero-field quasiparticle energy $E^{(0)}$ 
is shifted to $E=E^{(0)}\pm h_{\mbox{\small exc}}$. Therefore, the 
position of zero-energy states is now split 
to $\pm h_{\mbox{\small exc}}$.   
Figure~\ref{FIG:LDOS} plots the LDOS on the site nearest-neighboring to 
the impurity site. As is shown, when $h_{\mbox{\small exc}}=0$, there 
occurs a single ZEP in the LDOS. In the presence of exchange field, the 
ZEP is split into double peaks, each corresponding to one spin component. 
The magnitude of the splitting increases with the exchange field.
Since the local differential tunneling conductance is proportional to 
$\rho_{i}$, the ZEP and its splitting can be detected by the scanning 
tunneling spectroscopy (STS) as a test of the $d$-wave symmetry as 
well as the pre-existing exchange field in the above system. 
Experimentally, the nonmagnetic impurity with strong scattering potential 
can be realized by substitution of Zn for Cu in the CuO$_2$ layers. 
In addition, the STS best suited for exploring the local electronic 
properties allows a direct examination of whether the SC in 
Ru-1212 appears as a bulk state or can only survive at the boundaries 
between ferromagnetic domains. 
For the former, the STS data should reveal 
a superconducting gap over the whole sample.

Finally, we would like to mention once again that the bulk SC in Ru-1212 
prevails only when the exchange interaction is weak. For large exchange 
interaction, the SC could only exist at the interfacial CuO$_2$ layers 
between ferromagnetic domains, where $h_{\mbox{\small exc}}$ is small. 
Whether the SC in this compound is bulk or interfacial like appears to 
depend on sample 
preparation and this issue needs further experimental studies.  

{\em Note added:} After the paper was submitted for publication, we 
noticed a preprint by Shimahara and Hata~\cite{SS00} in which   
the enhancement of the possible FFLO state in a layered ferromagnetic 
compound such as Ru-1212 by the next-nearest-neighbor hopping was 
discussed.   

We are grateful to W. Kim, W. E. Pickett, Y. Y. Xue, and Kun Yang for 
valuable discussions. This work was supported by the Texas Center for 
Superconductivity at UH, a grant from the Robert A. Welch Foundation, 
the ARP-003652-0241-1999, the NSF at UH, and by the DOE at LBNL.

\begin{figure}
\caption{The $T$-$h_{\mbox{exc}}$ phase diagram for a $d$-wave 
superconductor with $\mu=-t$ (a), $-0.5t$ (b), and $-0.14t$ (c). The solid 
line represents the physical phase boundary between the superconducting 
state and the normal state. In panels (a) and (b), the dashed line 
indicates the transition between the FFLO state and the normal pairing 
(${\bf q}=0$) state. 
Note that the FFLO state exists only as the temperature is below 
the tri-critical point A. In panel (c), only  the transition 
between the normal pairing state and the normal state is allowed. 
} 
\label{FIG:PHASE} 
\end{figure}

\begin{figure}
\caption[h]{Magnetization as a function of the external magnetic field. 
The parameter values: $H_{c1}=100\;\mbox{G}$, $H_{int}=707\;\mbox{G}$, 
and $\kappa=50$. The inset shows the measured magnetization~\cite{Chu99}.}
\label{FIG:MAG} 
\end{figure}

\begin{figure}
\caption{Local density of states as a function of energy for 
$h_{\mbox{\small exc}}=0$ (solid line), $0.1\Delta_{0}$ (dashed line), 
$0.2\Delta_{0}$ (dotted line) at the site one lattice constant away from 
the impurity site. 
} 
\label{FIG:LDOS}
\end{figure}

\end{document}